\documentclass[%
 reprint,
 amsmath,amssymb,
 aps,
 pra,
 floatfix,
]{revtex4-1}

\newcommand{\ket}[1]{|#1\rangle}
\newcommand{\bra}[1]{\langle#1|}

\newcommand{\ketbra}[2]{\ket{#1}\bra{#2}}

\newcommand{\us}{\mathrm{\mu s}}

\usepackage{graphicx}
\usepackage{dcolumn}
\usepackage{bm}

\usepackage[urlcolor=blue, hyperindex, colorlinks, bookmarks=true]{hyperref}

\usepackage[utf8]{inputenc}

\usepackage{microtype}

\begin{document}

\title{Demonstration of a Parametrically-Activated Entangling Gate Protected from Flux Noise}

\author{Sabrina S. Hong}
\thanks{S. Hong, A. Papageorge, and P. Sivarajah contributed equally to this work.}
\author{Alexander T. Papageorge}
\thanks{S. Hong, A. Papageorge, and P. Sivarajah contributed equally to this work.}
\author{Prasahnt Sivarajah}
\thanks{S. Hong, A. Papageorge, and P. Sivarajah contributed equally to this work.}
\author{Genya Crossman}
\author{Nicolas Didier}
\author{Anthony M. Polloreno}
\author{Eyob A. Sete}
\author{Stefan W. Turkowski}
\author{Marcus P. da Silva}
\author{Blake R. Johnson}
\email[Corresponding author: ]{blake.johnson@ibm.com}
\affiliation{Rigetti Computing, 2919 Seventh St, Berkeley CA 94701}

\date{August 23, 2019}

\begin{abstract}

 In state-of-the-art quantum computing platforms, including superconducting qubits and trapped ions, imperfections in the 2-qubit entangling gates are the dominant contributions of error to system-wide performance. Recently, a novel 2-qubit \emph{parametric gate} was proposed and demonstrated with superconducting transmon qubits. This gate is activated through RF modulation of the transmon frequency and can be operated at an amplitude where the performance is first-order insensitive to flux-noise. In this work we experimentally validate the existence of this \emph{AC sweet spot} and demonstrate its dependence on white noise power from room temperature electronics. With these factors in place, we measure coherence-limited entangling-gate fidelities as high as $99.2\pm 0.15\%$.

\end{abstract}

\maketitle

There has been significant progress in recent years scaling up quantum processors, with several implementations demonstrating 20 or more qubits at various levels of maturity~\cite{google_ai_blog, IBMQ, Intel, Otterbach:2017}. However, in order to take advantage of increasing numbers of qubits, the limiting error rates of the devices must also improve commensurately. This is particularly true of near-term processors hoping to find utility in the so-called noisy intermediate-scale quantum (NISQ) regime that operate without the benefit of quantum error correction~\cite{Preskill:2018}. Errors in 2-qubit gates are the greatest impediment to deriving utility from today's superconducting quantum processors. The existing approaches to generating entanglement exhibit distinct trade-offs. For example, fast gates using flux-tunable qubits typically suffer from low coherence times anywhere except at a first-order flux-insensitive point (a so-called DC flux `sweet-spot')~\cite{Vion:2002}. Architectures that rely on flux tunability to bring qubits in and out of resonance must engineer high interaction rates and fast flux pulses in order to take only brief excursions away from the sweet-spot~\cite{DiCarlo:2009, Barends:2014, Rol:2019}. Alternatively, microwave activated gates, such as the cross resonance gate~\cite{Rigetti:2010, Chow:2011}, avoid flux tunability and the problems associated with it at the cost of typically slower interaction rates.

A recently-characterized technique for generating entanglement between a pair of capacitively-coupled transmons~\cite{Koch:2007} involves modulating the flux bias of a tunable transmon (i.e., parametrically modulating the qubit's frequency) in such a way as to drive a multi-transmon resonance~\cite{Beaudoin:2012, Strand:2013, McKay:2016, Naik:2017, Didier:2018, BluePaper, WhitePaper}---and for this reason this family of operations are referred to as \emph{parametric gates}. The stimulated Rabi process that results can, for instance, selectively drive population between the $\ket{11}$ and $\ket{20}$ states of the two-transmon system. A geometric phase accumulates as population undergoes a cycle in this two-level subspace, entangling the qubits. If applied for an appropriately chosen time, this interaction can be used to implement a controlled-Z (CZ) operation~\cite{Barenco:1995}.

One of the attractive features of these parametric gates is that the gate can be operated while remaining, on average, at the DC flux bias sweet-spot. However, excursions away from the sweet-spot during AC modulation may degrade dephasing times due to sensitivity to noise in both the DC bias signal as well as the AC modulation amplitude. Recent theoretical analysis of this problem has indicated that first-order insensitivity to flux noise can be recovered by operating the parametric gates at appropriately chosen amplitudes, leading to a novel operating point dubbed the \emph{AC sweet-spot}~\cite{Didier:2018b}. Since we operate the gate at an extremal value of the qubit frequency with respect to flux, the \emph{average} qubit frequency depends on the driven modulation amplitude. An asymmetric transmon has both a maximum and minimum frequency, leading to an AC sweet spot where the qubit's average frequency is insensitive to modulation drive amplitude.

In this work we experimentally validate the predicted behavior of dephasing at the AC sweet-spot, and exploit the enhanced dephasing time to demonstrate high-fidelity CZ operations. We focus, in particular, on the white-noise dependence of dephasing while operating at the AC sweet-spot. Theory predicts that, without appropriately chosen filtration that depends on the modulation frequency, performance can be limited by white noise. Consequently, we demonstrate that replacing commercial electronics used in prior experiments~\cite{BluePaper, WhitePaper, Otterbach:2017} with a custom arbitrary waveform generator (AWG) with an improved white noise floor leads to better gate performance at the AC sweet-spot. Our demonstration matches state-of-the-art fidelity benchmarks in superconducting qubits~\cite{Barends:2014, Sheldon:2016, Rol:2019} and is largely limited by relaxation rates observed in the device.

\begin{figure}[t!]
    \includegraphics[width=\linewidth]{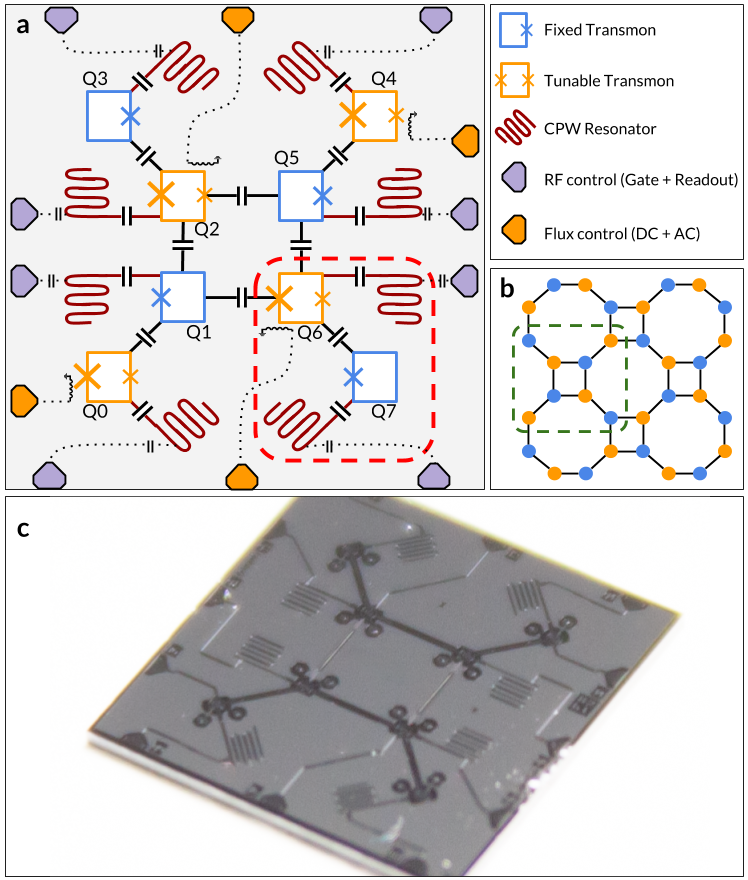}
    \caption{\textbf{Device diagram.} (a) Circuit diagram of the device under test. Our planar architecture features four frequency tunable transmons (orange) with a fixed capacitive coupling to four fixed frequency transmons (blue), each capacitively coupled to its own quarter-wavelength coplanar waveguide resonator (red) for readout. Single qubit control is implemented by driving microwave pulses through each qubit's resonator. Frequency tunable transmons each have their own inductively coupled flux bias line. All control lines (dotted lines) are in the same plane as circuit elements and external control is brought in from contact pads (purple for RF, orange for flux) at the edge of the chip, where individual wirebonds connect to a copper PCB. The two qubits used in this experiment (Q6, Q7) are denoted by the dashed red square. Coherence times and CZ fidelities of qubits and edges in the four-qubit subsystem defined by the neighbors of the tunable qubit are presented in Table~\ref{table:fidelities}. The coupling geometry and Hamiltonian of the chip represent a sub-component (dashed green) of the tileable lattice design shown in (b). (c) Shows a photograph of the fabricated device.}
    \label{fig:chipdiagram}
\end{figure}

To exploit the improved theoretical understanding of noise in parametric gates, we designed a multi-transmon device whose Hamiltonian allows for the operation of multiple gates at the AC sweet-spot. We then examine the coherence properties of transmons while driven with a modulating flux signal from two different AWGs and show that improved white-noise characteristics lead to full recovery of coherence times under modulation. This, in turn, enables high-fidelity entangling gates that we characterize in detail using randomized benchmarking~\cite{Magesan:2012}. The gates we study are selected from an eight-qubit device, wherein we focus on a tunable qubit coupled to three fixed-frequency neighbors, in the configuration shown in Fig.~\ref{fig:chipdiagram}.

\begin{figure}[t!]
    \includegraphics[width=\linewidth]{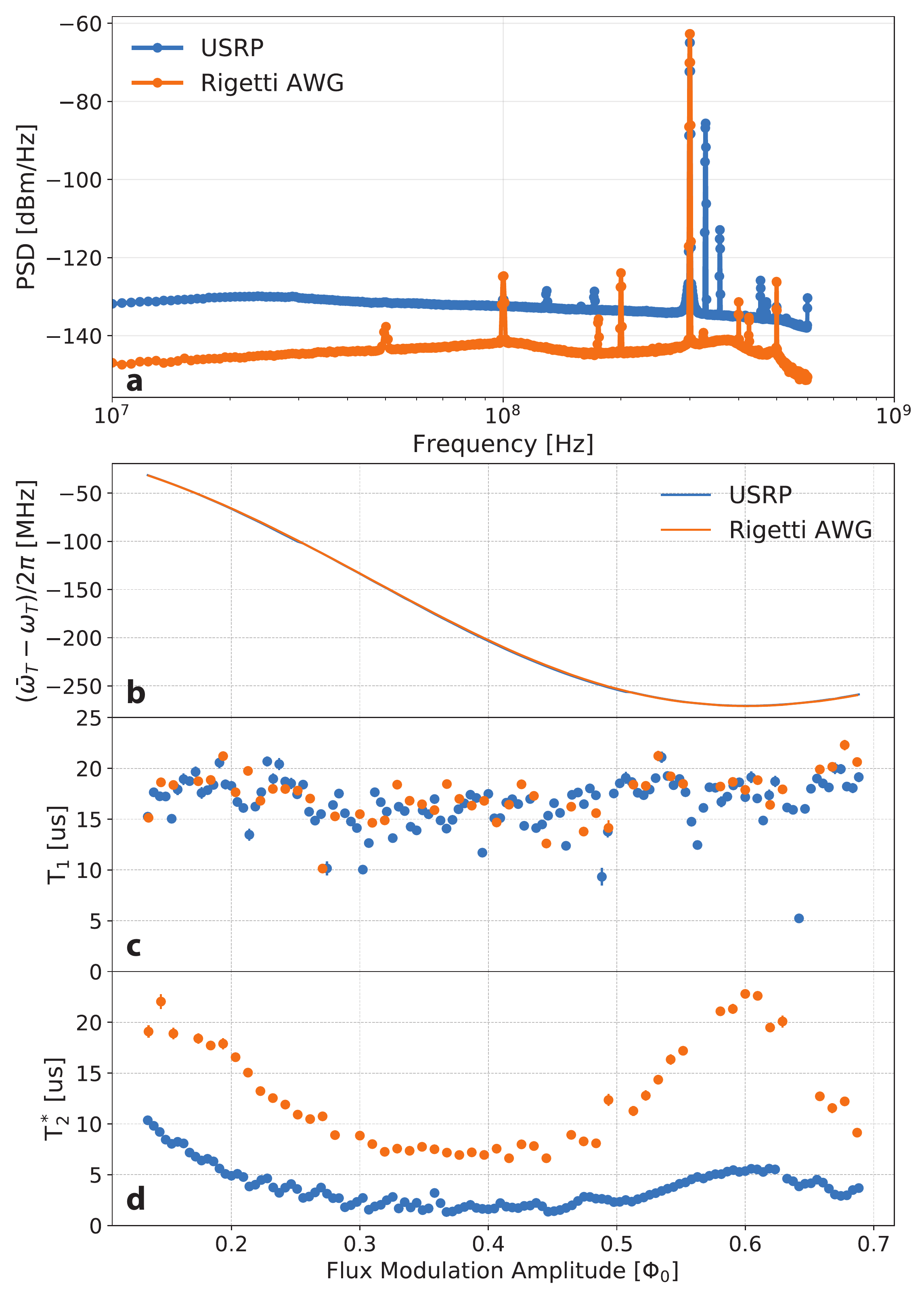}
    \caption{\textbf{Coherence under modulation.} We compare the coherence properties when modulating the flux with two different instruments: a National Instruments USRP (in blue) and a custom built flux delivery module (in orange). We present (a) the average power spectral density of each instrument measured with a pulsed output signal representative of typical gate parameters. We measure (b) the difference between the qubit's parking frequency and the time-averaged frequency under modulation as a function of the applied amplitude as well as (c)-(d) the coherence properties of the qubit under modulation. In both the $T_{1}$ and $T_{2}^{*}$ experiments in (c) and (d), we replace the free evolution time of the experiment with a time-varying modulated flux pulse at a fixed frequency. We do not expect a $T_{1}$ dependence from this modulation, but we use the measured relaxation to calculate the pure dephasing time $T_{\phi}$. We report all modulation amplitudes in units of the flux quantum by finding the linear scaling factor consistent with a minimum $\delta\omega_T$ at a modulation amplitude of $0.6\,\Phi_0$.}
    \label{fig:dum}
\end{figure}

A two-qubit subsystem of our device comprises one tunable and one fixed-frequency transmon with frequencies $\omega_T$ and $\omega_F$ and anharmonicities $\eta_T$ and $\eta_F$, respectively. The qubits are coupled via a fixed capacitive coupling at a rate $g$. The SQUID of the tunable transmon is asymmetric and thus exhibits DC sweet spots~\cite{Vion:2002} at two frequencies (at flux biases of 0 and $0.5\,\Phi_0$). It is coupled to a flux bias line that allows for DC and AC control of the transmon frequency. We choose the DC flux bias to operate the tunable transmon at its maximum frequency. The combined applied flux is $\Phi(t) = \Phi_\mathrm{dc} + \epsilon(t) \cos(\omega_p t)$, where $\epsilon(t)$ is the envelope of a carrier at frequency $\omega_p$, resulting in time-dependent modulation of $\omega_T(t) \approx \bar\omega_T(\epsilon) + \lambda(\epsilon)\cos(2\omega_p t)$, where $\lambda(\epsilon)$ is the amplitude-dependent conversion factor between flux and frequency. Thus, modulating $\omega_T(t)$ around its maximum value has the dual effect of offsetting the average qubit frequency by $\delta\omega_T = \bar\omega_T - \omega_T^\mathrm{max}$ (Fig.~\ref{fig:dum}b), and generating sidebands at even harmonics of the modulation frequency. By appropriate choice of modulation frequency and amplitude, these sidebands may be used to drive resonant interactions involving states of the multi-qubit system.

Dropping terms off-resonant from the flux drive~\cite{Didier:2018}, the interaction Hamiltonian when modulating at one of the resonance conditions is
\begin{align}
    \hat{H}_\mathrm{int} = & g_\mathrm{eff} \exp \left\{i\left(2\omega_p - |\Delta + \eta_T + \delta\omega_T|\right)t \right\} \ketbra{11}{02} \notag\\
                           & + \mathrm{h.c.},
\end{align}
where $\Delta = \omega_T^\mathrm{max} - \omega_F$ is the static qubit-qubit detuning and $g_\mathrm{eff}$ is the effective coupling rate. At small modulation amplitudes, $g_\mathrm{eff} \approx \sqrt{2}\hbar gJ_1\left(\frac{\delta\omega_T}{2\omega_p}\right)$, where $J_1$ is the Bessel function of the first kind~\cite{Didier:2018}. In terms of qubit spectroscopy, the $\ket{11}$ and $\ket{20}$ states are directly coupled, and population is exchanged between them when the resonance condition $|\Delta + \eta_T + \delta\omega_T | = 2\omega_p$ is satisfied. Since $\omega_T(t)$ depends on the amplitude of the flux modulation, the resonance exists as a contour in amplitude-frequency space. This contour exhibits a point of vanishing derivative with respect to flux amplitude at $\epsilon(t) \approx 0.6\,\Phi_0$ (see Fig.~\ref{fig:dum}b)---this is the AC sweet spot. In addition, operating the gate in this fashion reduces sensitivity to fluctuations and drift in the amplitude of the flux modulation.

The qubit's flux bias is subject to noise in the DC offset as well as in the AC amplitude of the flux modulation driving the parametric transition. In general, both of these noise sources lead to dephasing, which, in turn, degrades the gate performance. This is an issue that all flux-tunable qubits must contend with given the universal nature of $1/f$ flux noise and ongoing research into its microscopic origin~\cite{Wellstood:1987, Choi:2009, Wang:2015, Kumar:2016}. Under modulation around a DC sweet spot~\cite{Vion:2002}, however, $1/f$ noise on the DC offset is dynamically decoupled. Remarkably, for specific modulation amplitudes where the average frequency is flat, the qubit also becomes first-order insensitive to $1/f$ noise on the AC amplitude. In analogy to the DC sweet spots of flux-tunable superconducting qubits~\cite{Vion:2002}, we call these operating points AC sweet spots~\cite{Didier:2018b}. Other flux-tunable gate schemes have no protection from $1/f$ noise, and thus rely on short gate duration to achieve high fidelity~\cite{DiCarlo:2009, Barends:2014, McKay:2016}.

The physical device used in our experiment is a superconducting aluminum circuit fabricated on a high-resistivity silicon wafer. The eight-qubit device consists of four tunable and four fixed-frequency transmon qubits, each capacitively coupled to its own quarter-wavelength coplanar waveguide resonator as shown in Fig.~\ref{fig:chipdiagram}a. The transmons are coupled along the edges of a truncated square tiling lattice (i.e., a lattice with vertex configuration 4.8.8), with an alternating arrangement of fixed and tunable transmons at adjacent lattice sites. Single-qubit control is implemented by driving microwave pulses through each qubit's resonator at the qubit's $\ket{0}\rightarrow\ket{1}$ transition frequency, while state interrogation is implemented by driving at the resonator's frequency~\cite{Blais:2004}. State preparation is achieved by waiting several multiples of $T_1$ between experimental cycles. The flux control lines are inductively coupled to the SQUID loops of the tunable qubits. A partial schematic of the experimental setup is shown in the appendix Fig.~\ref{fig:fridgediagram}. We focus on the qubit pair Q6-Q7; Q6 is a triply-connected tunable transmon and representative of an interaction that may be achieved in a larger lattice. The maximum and minimum frequencies of Q6 are 4.475~GHz and 4.080~GHz, respectively, while Q7 has a fixed frequency of 3.826~GHz. Both qubits have an  anharmonicity of $\sim$200~MHz. Other device parameters are listed in Table~\ref{table:fidelities}.

In order to assess the impact of instrumentation noise on the performance of the parametric gate, we establish baseline noise measurements using two separate arbitrary waveform generators (AWGs). The first is a National Instruments USRP X300 software-defined radio with UBX160 daughter boards, which was used in prior parametric gate demonstrations~\cite{BluePaper, WhitePaper, Otterbach:2017}. The second is a custom Rigetti AWG designed particularly for this application. Using pulse parameters that are representative of those used to drive parametric resonances, we measure a power spectral density from a train of identical pulses generated by each AWG, as illustrated in Fig.~\ref{fig:dum}a. Away from the 300~MHz signal peak, we see as much as 15~dBm/Hz reduction in the noise power with the Rigetti AWG, and improved spur performance across the entire band with the exception of subharmonics at 100 and 200~MHz. The custom Rigetti AWG has significantly less white noise because it generates analog signals directly through digital synthesis, as apposed to the USRP's mixer-based architecture. Both instruments employ low noise digital-to-analog converters and amplifiers, but the mixers on the USRP add significant white noise. Furthermore, the Rigetti instrument employs a push-pull amplifier front-end to suppress even harmonics of the output signal frequency.

We then use both instruments to measure the coherence time under modulation, a critical parameter for gate performance, by performing $T_1$ and Ramsey experiments interspersed by flux pulses produced by either instrument, Fig.~\ref{fig:dum}c-d. We measure coherence time as a function of the amplitude of the flux modulation, and observe the AC sweet spot predicted in Ref.~\cite{Didier:2018}, as evidenced by a resurgence in $T_2^*$ using either instrument in Fig.~\ref{fig:dum}d. Not only does the Rigetti AWG demonstrate uniformly better coherence properties, but also the coherence at the AC sweet spot matches the coherence at 0-flux amplitude, a property we refer to as \emph{full resurgence}. This is in contrast to the resurgence effected by the USRP, which is 60\% with respect to the $T_2^*$ measured at 0-flux amplitude. The marked difference in $T_2^*$ is attributed to the difference in white-noise floors between the two signal generators used to produce the flux modulation. The results shown in the remainder of the text use the Rigetti AWG to drive flux pulses.

\begin{figure}[t!]
    \includegraphics[width=\linewidth]{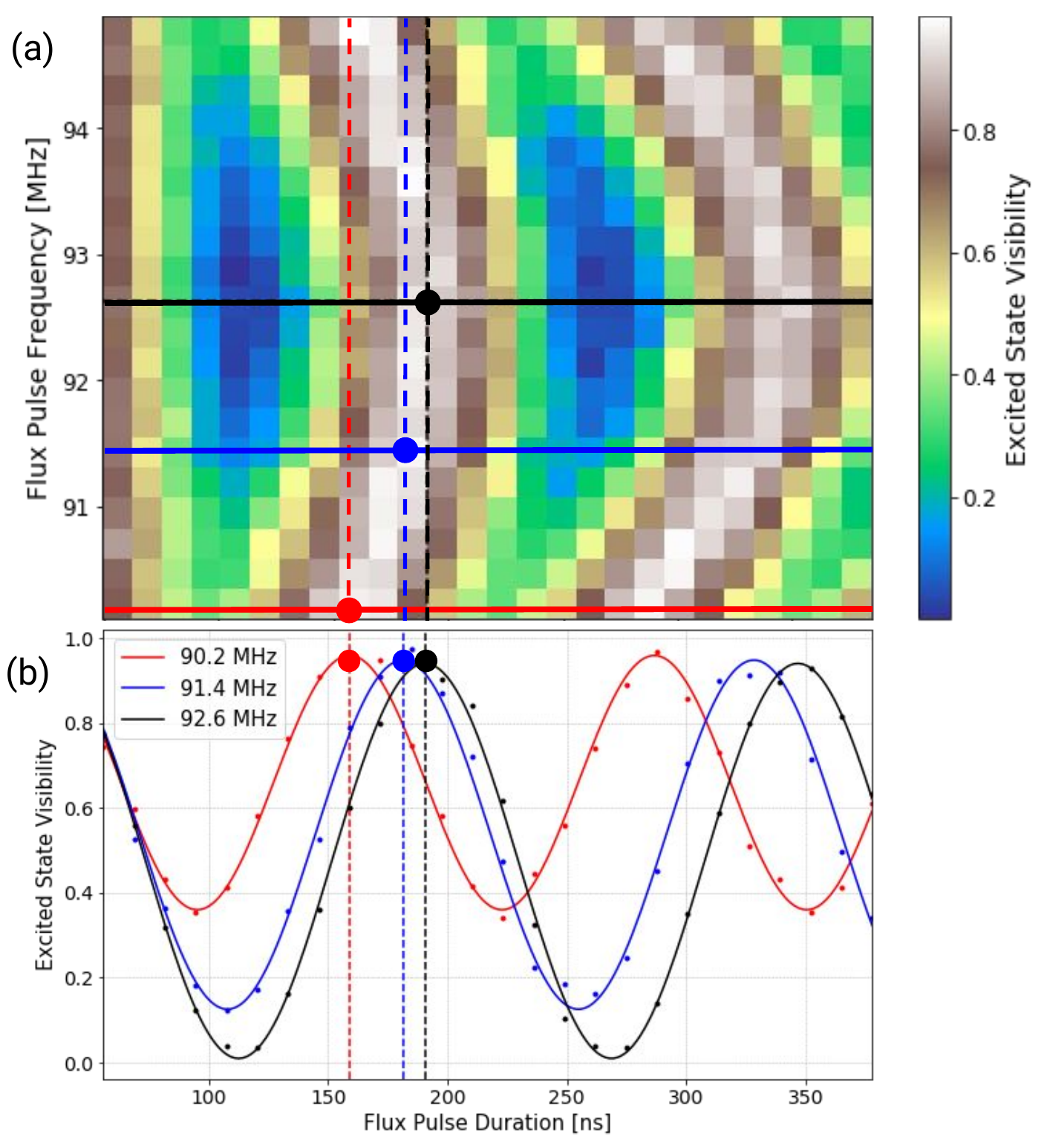}
    \caption{\textbf{Parametric Gate Calibration.} (a) Excited state population of the fixed qubit, Q7, when preparing the $\ket{11}$ (fixed-tunable) state and driving the $\ket{11} \leftrightarrow \ket{02}$ transition near the AC sweet spot. This results in a characteristic ``chevron'' pattern as we vary the frequency and duration of the flux pulse. To calibrate the gate, we fit slices of fixed pulse frequency and varying duration, like those shown in (b), to find the pulse duration that most closely returns the system to the $\ket{11}$ state.}
    \label{fig:chevron}
\end{figure}

To enact the parametric CZ gate, a single-frequency flux pulse with an envelope defined by a constant section and symmetric error function shoulders is calibrated over its amplitude, duration, and frequency to maximally entangle the two qubits~\cite{WhitePaper, BluePaper}. This is accomplished by first identifying the resonant frequency that corresponds to operating the gate at an AC sweet spot, and then empirically determining the amplitude where $d\bar\omega_T/d\epsilon$ is small, as shown in Fig.~\ref{fig:dum}b. Using this amplitude, the frequency is tuned by first preparing the qubits in the $\ket{11}$ state and optimizing population transfer to the $\ket{02}$ (fixed-tunable) state with maximal visibility. The parametric gate may be understood as nutation in the $\ket{11}-\ket{02}$ subspace, where the geometric phase accrued in this space corresponds to the entangling phase of an associated CPHASE($\phi$) = diag(1,1,1,$e^{i\phi}$) gate, so long as population is completely returned to the $\ket{11}$ state. By optimizing the frequency and duration of the flux pulse we can freely choose the entangling phase, $\phi$. In practice, we measure the interaction as a function of pulse frequency and duration near resonance, then fit to a cosine model to find the appropriate gate time for each frequency, as shown in Fig.~\ref{fig:chevron}. Ramsey experiments are used to extract the entangling phase as well as the single-qubit Z rotations, and an operating point is selected which most closely enacts CPHASE($\pi$), \emph{i.e.} CZ. At this point we can directly extract the effective coupling rate, $g_\mathrm{eff}/2\pi \approx 3.4\,\mathrm{MHz}$. The resulting flux pulse is 176~ns long with 24~ns rise and fall, and modulated at 92~MHz. Note that Z rotations may be absorbed into the single-qubit control frames, and so they merely need to be calibrated and used in the resulting gate definition.

To assess the quality of the resulting CZ gate, we repeatedly perform interleaved randomized benchmarking (iRB)~\cite{Magesan:2012} over approximately eight hours. Each iRB experiment comprises a collection of `reference' sequences drawn from the two-qubit Clifford group, and a collection of `interleaved' sequences wherein a particular gate is interspersed between each random Clifford gate. Fluctuations in the coherence times (notably $T_1$~\cite{Gustavsson:2016,Klimov:2018}) over the duration of an iRB experiment can result in incorrect estimates of the fidelity. In particular, because iRB compares the reference sequences to the interleaved sequences to infer the fidelity of the gate under study, any difference in the decoherence rate will be ascribed to properties of that gate. If these experiments are performed sequentially and the decoherence rate varies temporally, the estimate of the fidelity can be too high or too low, depending on the direction of the temporal variation.

\begin{figure}[tb!]
    \includegraphics[width=\linewidth]{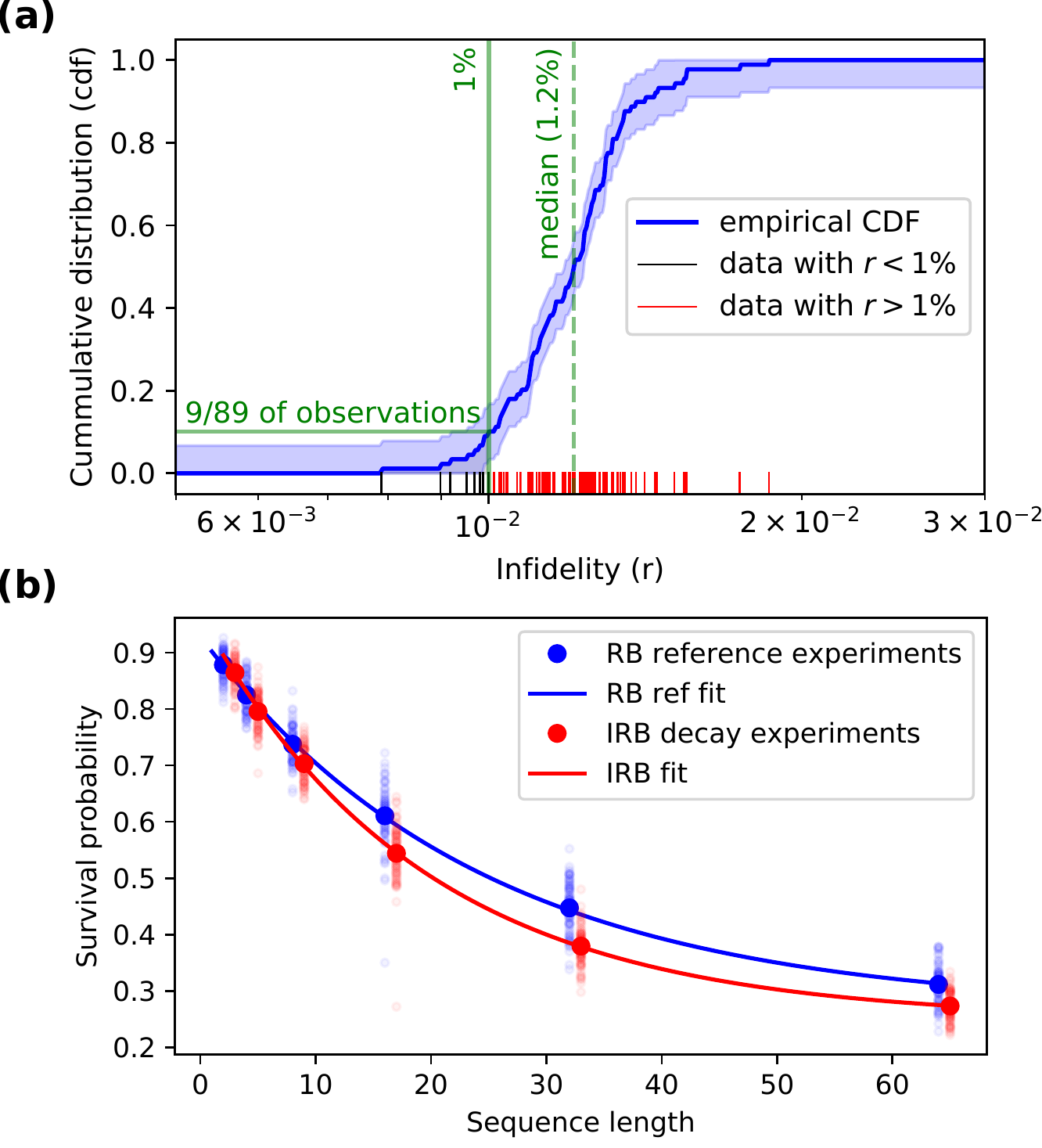}
    \caption{\textbf{Repeated benchmarking.} (a) We estimate the empirical cumulative distribution function (ECDF) from repeated observations of gate infidelity as measured by interleaved randomized benchmarking over a period of 8 hours (90\% confidence region for the ECDF shown). We report a gate fidelity $>99\%$ for $\sim$10\% of the recorded fidelities, with the highest recorded fidelity at $99.2\pm 0.15\%$ over this period shown in (b). For these RB sequences, we extract decay rates of $p_\mathrm{ref} = 0.960 \pm 0.086$ and $p_\mathrm{int} = 0.950 \pm 0.091$, from which we can estimate the mean (across all 11,520 different 2-qubit Clifford group operations~\cite{Corcoles:2013}) average gate infidelity to be $\approx 97\%$.
}
    \label{fig:ecdf}
\end{figure}

To account for this we modified the iRB protocol such that we could test whether the behavior of the experiments changed appreciably. Instead of measuring a reference RB decay followed by an interleaved RB decay, we grouped experiments to measure two reference decays and two interleaved decays. Moreover, we scrambled the order in which data was collected among the 4 RB experiments of each group to remove any effective temporal order between them. We then performed bootstrap hypothesis testing~\cite{Boostrap} to determine if it was possible to distinguish between the two reference RB decays (and similarly for the two interleaved RB decays), discarding experiments that gave even weak evidence that the decay rates were different (the significance level was set at $10\%$). For the experiments that remained, we combined the two copies of each decay, and computed the point iRB estimate of the average gate infidelity~\footnote{See eqn.~4 in Ref.~\cite{Magesan:2012}.}. Of the 102 experiments, only 13 were discarded. The resulting distribution of point iRB estimates is shown in Fig.~\ref{fig:ecdf}a. Uncertainties for individual iRB estimates of average fidelity were less than $\pm 0.4\%$~\footnote{All uncertainties quotes are 90\% confidence intervals. See Appendix~\ref{sup:irb} for details of the fit analysis.}. All experiments had infidelities below 2$\%$, and 9 of the 89 post-selected experiments had infidelities below 1$\%$. The best observed fidelity was $99.19\pm0.15\%$ (an infidelity of $0.81\pm0.15\%$)~\footnote{The quoted results are point estimates from iRB, assuming all errors are depolarizing. A rigorous treatment of the inversion problem in iRB leads to region estimates, such as eqn.~5 of Ref.~\cite{Magesan:2012} or eqn.~6.1 of Ref.~\cite{Kimmel:2014} with much coarser resolution. Using the region estimate from Ref.~\cite{Kimmel:2014}, the best point estimate yields the region 88.21\%--99.88\%, while the ensemble of post-selected experiments yields regions with lower bound in the range 84.58\%--88.21\%, and upper bounds in the range 99.79\%--99.95\%. For region estimates we ignore statistical uncertainty since those corrections are nearly two orders of magnitude smaller.}, with corresponding iRB decays shown in Fig.~\ref{fig:ecdf}b~\footnote{The distribution does not change appreciably if we do not post-select experiments via the hypotheses test for stability. It does change the number of experiments below 1$\%$ infidelity slightly, but all experiments remain below 2$\%$ infidelity.}.

Along with the iRB experiments we also measured coherence times ($T_1$ and $T_2^*$) under modulation and attempt to corroborate the observed gate fidelity to a static model with time-independent decoherence rates~\footnote{See Appendix~\ref{sup:irb}}. Considering the coherence times in aggregate, with $T_1$ under modulation measuring 10.5--20.3~$\mu s$ and 18.1--29.9~$\us$ and $T_2^*$ under modulation measuring 10.5--18.0~$\us$ and 16.4--21.8~$\us$ for the fixed and tunable qubits respectively, the aggregate prediction for the average gate fidelity is 97.6--98.7\%. Consequently, the observed distribution of infidelity is consistent with the variation in the coherence times.

In summary, we have demonstrated a high-fidelity, coherence-limited, parametrically-activated two-qubit gate on a multi-qubit architecture via direct modulation of the tunable qubit. The device in question was designed to work in a general purpose multi-qubit configuration---it is not a one-off design exploiting favorable features that cannot be reproduced in larger lattices. The parametric gate we study is highly selective and robust to crowded spectroscopy, in the sense that one may operate high-fidelity gates so long as the relevant transitions are separated by $\sim 5g$, or approximately 25~MHz for the parameters of this device. This makes the parametric gate well-suited to enact pairwise entangling operations in large lattices. On this particular device, junction fabrication parameters on neighboring qubits yielded a frequency configuration with especially slow gates for other pairs in the lattice~\footnote{See Table~\ref{table:fidelities} for performance of other gates in the lattice}. Improvements in fabrication and robust Hamiltonian design will increase the yield of such devices, allowing for the scalable operation of multi-qubit devices with current infrastructure.

\begin{acknowledgments}

This work was funded by Rigetti \& Co Inc., dba Rigetti Computing. We thank Colm Ryan for critical reading and useful discussions, Amy Brown for developing CPHASE calibration tools, Joshua Combes for discussions about the statistical analysis, Matt Reagor for the initial Hamiltonian design to operate at the AC sweet spot, and the Rigetti fabrication team for manufacturing the device. This result relied heavily on the efforts of the Rigetti control systems and embedded software teams to create the Rigetti AWG control system.

A.T.P., M.P.S. and B.R.J. drafted the manuscript. S.S.H, A.T.P., and P.S. performed the experiment. G.C. designed the chip. N.D. and E.A.S. provided theory of the gate's operation and dephasing under modulation. A.M.P. did additional gate characterization. S.W.T designed the low-noise analog front-end of the Rigetti AWG. M.P.S. designed and carried out the analysis procedure. B.R.J. organized the effort and finalized the manuscript.

\end{acknowledgments}

\bibliography{references}

\appendix

\section{Experimental Setup}
\label{sup:setup}
\begin{figure}[tb!]
    \includegraphics[scale=.5]{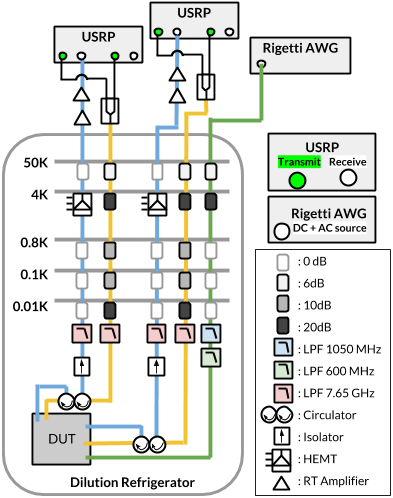}
    \caption{\textbf{Overview of the experimental setup.} This diagram details the control electronics, wiring, and filtering for the two qubits involved in the experiment (Q6 and Q7). The single qubit control pulses and readout pulses for one qubit are generated separately on two daughter cards of one USRP software defined radio. These signals are combined at room temperature and sent down one line in the dilution refrigerator (yellow). The readout signal (blue) is first amplified by a high electron mobility transistor (HEMT), followed by two room temperature amplifiers before being received by the USRP receive port. Both DC and AC signals for flux delivery (green) are generated by custom Rigetti AWGs. All control lines go through various stages of attenuation and filtering in the dilution refrigerator.}
    \label{fig:fridgediagram}
\end{figure}

The physical device used in our experiment is packaged and mounted in a dilution refrigerator and cooled to 10mK. The sample is mounted to a copper PCB using 1\% Si/Al wirebonds and packaged in a light-tight assembly through which DC and microwave signals are delivered via non-magnetic SMPM surface mount connectors. An overview of the experimental setup used to address the two qubits used in this experiment is shown in Figure~\ref{fig:fridgediagram} where each individual component is addressed. Note that the actual state of the system during the experiments also included similar setups for all other qubits on the eight-qubit device under test (four tunable and four fixed).

\section{Chip Performance}
\label{sup:performance}
We characterize an eight-qubit device where the experiment is performed on a pair of qubits where the tunable qubit of interest is itself coupled to three fixed qubits. Measured coherence times and CZ fidelities of qubits and edges in this four-qubit system are presented in Table~\ref{table:fidelities}. We focus on the highest performing pair Q6 and Q7 for the detailed experiments and analysis.

\begin{table*}[tb!]
\setlength{\tabcolsep}{8pt}
\begin{tabular}{l*{7}{c}}
 \hline\hline
  \rule{0pt}{13pt}
  & $\mathcal{\overline{F}_\mathrm{CZ}}$ (\%)
  & $t_\mathrm{gate}$ (ns)
  & $\omega_p/2\pi$ (MHz)
  & $\widetilde{T}_1^F$ ($\mu s$)
  & $\widetilde{T}_1^T$ ($\mu s$)
  & $\widetilde{T}_2^{*F}$ ($\mu s$)
  & $\widetilde{T}_2^{*T}$ ($\mu s$)\\
 \hline
 Q6-Q7 & 98.8 & 176 & 92  & 13.9--17.9 & 21.9--25.3 & 13.5--15.8 &  18.2--20.7  \\
 Q6-Q1 & 97.4 & 292 & 125 &   26.4    &    18.9    &   10.4    &    25.9 \\
 Q6-Q5 & 94.5 & 336 & 185 &   31.4    &    20.9    &   21.4    &    18.0 \\
 \hline\hline
\end{tabular}
\caption{\textbf{Performance results.} Gate performance for the multi-qubit unit cell defined by the tunable qubit, Q6. We calibrated CZ parametric gates on all pairs of the chip and present results from the highest performing pair (Q6-Q7), as well more cursory analysis of other pairs connected to Q6. We show the average CZ fidelity as measured by iRB, the corresponding gate duration, $t_\mathrm{gate}$, and the coherence times under modulation at the amplitude and frequency of the corresponding gate: $\widetilde{T}_1^F$, $\widetilde{T}_1^T$, $\widetilde{T}_2^{*F}$, and $\widetilde{T}_2^{*T}$. The pair Q6-Q7 received significantly greater scrutiny, so in this case we show the interquartile ranges of $\widetilde{T}_1$ and $\widetilde{T}_2^*$ over 211 measurements of these quantities that were interspersed with the iRB experiments. Note that while the tunable qubit is common to all the coherence numbers in the table, the modulation conditions under which the decay constants were probed are different for each pair.}
\label{table:fidelities}
\end{table*}

\section{2-qubit Gate Set Tomography (GST)}
\label{sup:gst}

\begin{figure}[htb!]
    \includegraphics[width=\linewidth]{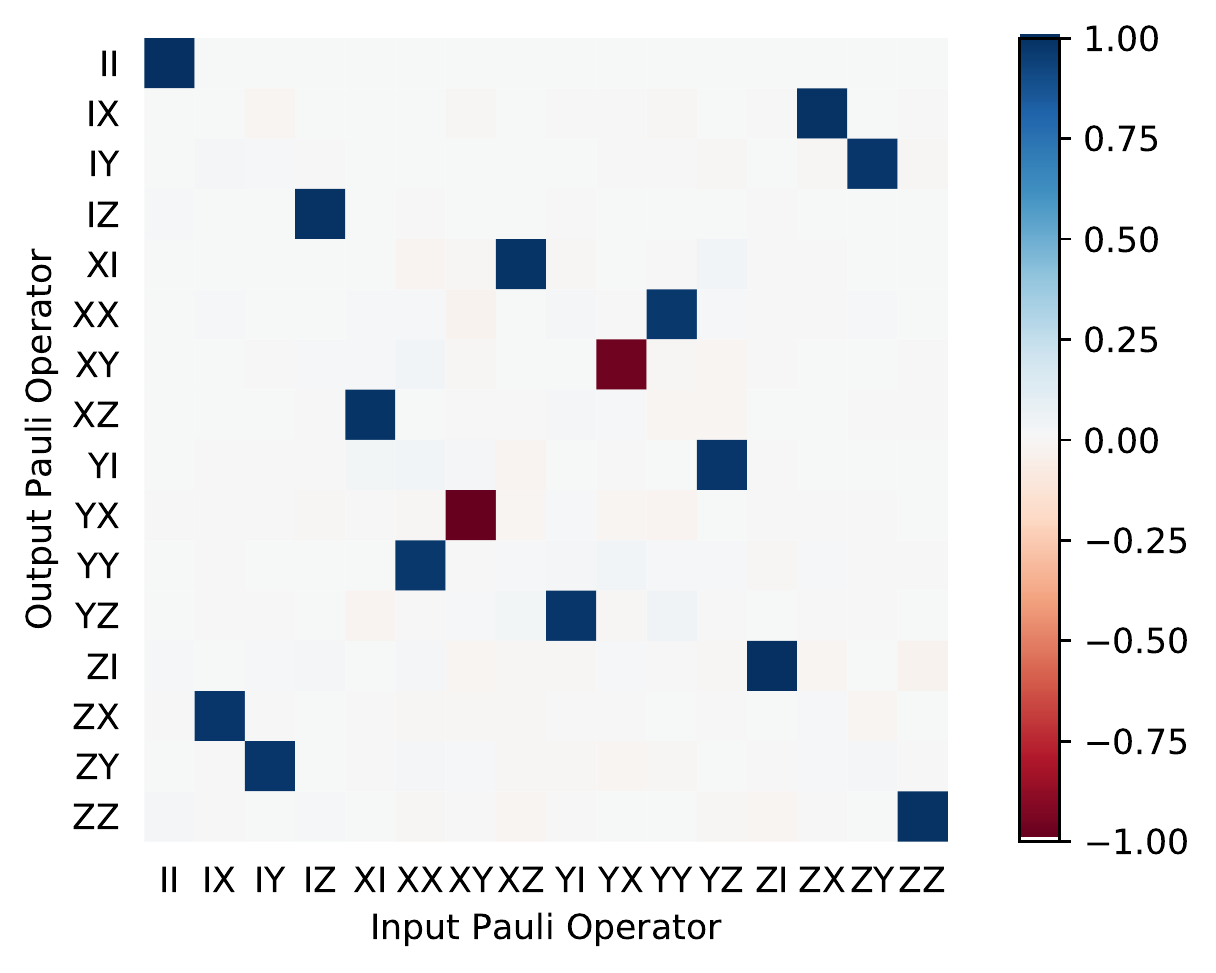}
    \caption{\textbf{Gate Set Tomography.} The reconstructed CZ Pauli transfer matrix}
    \label{fig:gst}
\end{figure}

As an independent validation of the gate performance, we also performed gate-set tomography (GST)~\cite{Blume-Kohout:2017},
using the pyGSTi library~\cite{pyGSTi}. Using GST we measure the average gate fidelity to be 98.2\%, which is consistent with
the point iRB estimates we obtained, and indicates that the region estimates for iRB are indeed pessimistic.
Fig.~\ref{fig:gst} shows a representative tomogram acquired using GST.

Our GST experiment were performed in 13 minutes, using a reduced number of fiducial pairs and sequence lengths of 1, 2, 4, 8,
16, and 32 -- in contrast, our iRB experiments were performed in approximately 4 minutes. The shortcoming of GST
is in the sheer
number of sequences that must be performed to self-consistently produce an estimate of the multi-qubit parameters,
making it sensitive to temporal variation of decoherence rates over these time scales. A log-likelihood ratio
goodness of fit test for the overall GST fit for a time-independent Makorvian model was 71 standard deviations away
from the expected value of the log-likelihood ratio statistic, indicating that temporal variation of decoherence rates
(or other model violations) are of high statistical significance. In future work we will investigate how to improve the GST
fits.

\section{Data analysis for repeated iRB, T1, T2}
\label{sup:irb}

\begin{figure*}[htb!]
    \centering
    \includegraphics[width=\textwidth]{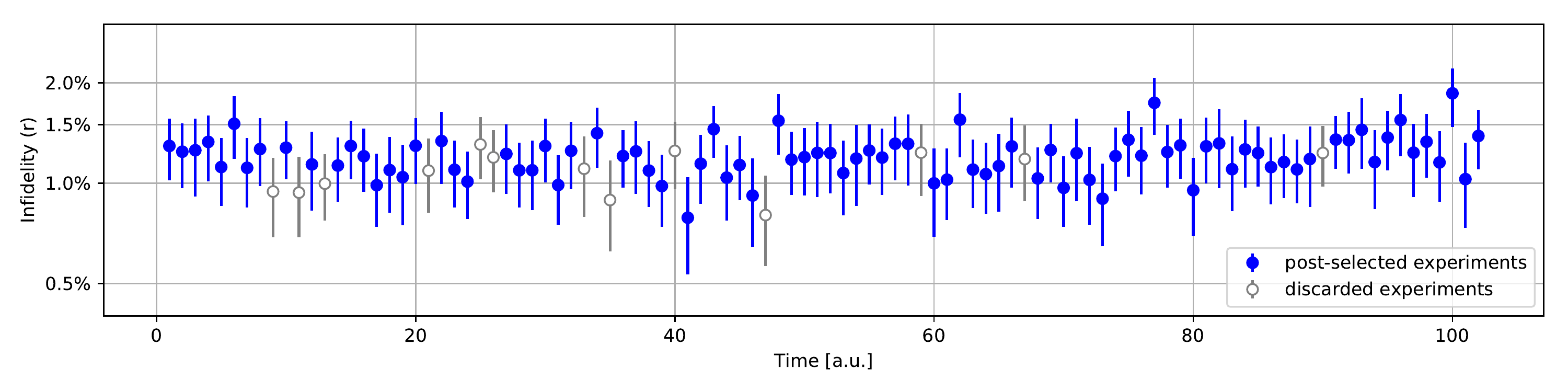}
    \caption{Time-series for iRB infidelity for CZ.  Error bars correspond to 90\% confidence intervals, and gray/empty data points corresponds to iRB experiments that were excluded due to failing the stability hypothesis test (i.e., decay rates were not stable for the duration of the experiments). The time span for the measurement corresponds to 8h.}
    \label{fig:irb-time-series}
\end{figure*}

To validate our result against changing decoherence rates (notably $T_1$) over time, we perform repeated measurements of iRB over the course of eight hours. For each reference and interleaved decay of survival probability, random sequences of Clifford group operations of lengths 2, 4, 8, 16, 32, and 64 were generated, and each Clifford group operation was compiled into sequences of single qubit rotations and CZ gates (for the interleaved, each Clifford was decomposed into a product of CZ and another Clifford). The survival probability fits were performed using a weighted non-linear least squares estimator for the model $A p^L + B$~\cite{Magesan:2011}, where weights were based on the inverse variance of survival probabilities across random sequences for a fixed length. Each random sequence was measured 500 times, and 32 random sequences were generated per length.

The 90\% confidence intervals are generated by a parametric percentile bootstrap, where the counts for the 500 measurements of each fixed random sequence were resampled for a binomial distribution with 500 samples and $p$ equal to the sample mean. A total of 2000 bootstrap replicants were generated for each set of experiments.

As described in the main text, each iRB estimate consists of two reference RB experiments, two interleaved RB experiments, a $T_1$ experiment, and a $T_2^*$ experiment. Before running these experiments, we enumerated all sequences to be measured for these 4 classes of experiments, and then randomized the order in which the sequences were measured, so that in effect there was no clear temporal ordering between the $T_1$, $T_2^*$, and iRB estimates (i.e., they were, in effect, measured simultaneously). We then applied bootstrap hypothesis testing to ensure each of $p$ estimates for the two reference RB decays were consistent (at a 10\% significance level), taking that to be an indication of $T_1$ fluctuating in time (which may bias the iRB estimate). We discarded sets of experiments where either the reference or the interleaved decays were not consistent, but find that this post-selection did not significantly change the distribution of iRB estimates.

The fidelities reported are \emph{average gate fidelities}~\cite{Bowdrey:2002,Nielsen:2002}, which are related to the RB $p$ via $\overline{F}=\frac{(d-1)p+1}{d}$~\cite{Carignan-Dugas:2016}, where $d$ is the system dimension ($d=4$ in our case, since we have 2 qubit gates). The infidelity $r$ is simply $r=1-\overline{F}$.

\end{document}